%% file: Tie_breaking_rule_collusion_v3.tex
\pdfoutput=1
\documentclass[12pt]{article}
\usepackage[utf8]{inputenc}
\usepackage[british]{babel}
\usepackage{cmap}
\usepackage{lmodern}

\usepackage{amssymb, amsmath, amsthm}
\usepackage[a4paper,top=25mm,bottom=25mm,left=25mm,right=25mm]{geometry}
\usepackage{ragged2e}

\usepackage{authblk} 
\usepackage{pifont}
\usepackage{graphicx}
\usepackage[dvipsnames,svgnames,table]{xcolor}
\usepackage[figuresright]{rotating}
\usepackage{xtab} 
\usepackage{longtable} 
\usepackage{multirow}
\usepackage{footnote}
\usepackage[stable]{footmisc}
\usepackage{chngpage} 
\usepackage{pdflscape} 
\usepackage{tocbibind} 

\usepackage{pgfplots}
\pgfplotsset{compat=1.18}
\pgfplotsset{every tick label/.append style={font=\footnotesize}}
\usepackage{setspace}

\makesavenoteenv{tabular}
\usepackage{tabularx}
\usepackage{booktabs}
\usepackage{threeparttable} 
\usepackage[referable]{threeparttablex} 
\newcolumntype{R}{>{\raggedleft\arraybackslash}X}
\newcolumntype{L}{>{\raggedright\arraybackslash}X}
\newcolumntype{C}{>{\centering\arraybackslash}X}
\newcolumntype{A}{>{\columncolor{gray!25}}C}
\newcolumntype{a}{>{\columncolor{gray!25}}c}

\newlength{\tablen}

\usepackage{dcolumn} 
\newcolumntype{.}{D{.}{.}{-1}}

\usepackage{tikz}
\usetikzlibrary{arrows, calc, matrix, patterns, positioning, trees}
\usepackage[semicolon]{natbib}
\usepackage[hyphens]{url}
\usepackage{hyperref} 
\hypersetup{
  colorlinks   = true,    		
  urlcolor     = blue,    		
  linkcolor    = blue,			
  citecolor    = ForestGreen,		  	
}
\usepackage{microtype}
\usepackage[justification=centering]{caption} 

\usepackage[labelformat=simple]{subcaption}

\DeclareCaptionLabelFormat{parenthesis}{(#2)}
\makeatletter
\renewcommand\p@subfigure{\arabic{figure}.}
\makeatother

\DeclareCaptionLabelFormat{parenthesis}{(#2)}
\makeatletter
\renewcommand\p@subtable{\arabic{table}.}
\makeatother

\def\addlegendimage{\csname pgfplots@addlegendimage\endcsname}

\usepackage{enumitem}

\setlist[itemize]{leftmargin=2.5\parindent}
\setlist[enumerate]{leftmargin=2.5\parindent}

\theoremstyle{plain}

\theoremstyle{definition}

\newtheorem{example}{Example}

\theoremstyle{remark}

\def\keywords{\vspace{.5em} 
{\noindent \textit{Keywords}: }}

\def\AMS{\vspace{.5em} 
{\noindent \textbf{\emph{MSC} class}: }}

\def\JEL{\vspace{.5em} 
{\noindent \textbf{\emph{JEL} classification number}: }}

\title{On head-to-head results as tie-breaker \\ and consequent opportunities for collusion}
\author{\href{https://sites.google.com/view/laszlocsato}{L\'aszl\'o Csat\'o}\thanks{~E-mail: \emph{laszlo.csato@sztaki.hu}} }
\affil{Institute for Computer Science and Control (SZTAKI) \\
Hungarian Research Network (HUN-REN) \\
Laboratory on Engineering and Management Intelligence \\
Research Group of Operations Research and Decision Systems}
\affil{Corvinus University of Budapest (BCE) \\
Institute of Operations and Decision Sciences \\
Department of Operations Research and Actuarial Sciences}
\affil{Budapest, Hungary}
\date{\today}

\def\Dedication{
{\noindent
$\mathfrak{Was}$ $\mathfrak{sich}$ $\mathfrak{hier}$ $\mathfrak{in}$ $\mathfrak{der}$ $\mathfrak{Entwicklung}$ $\mathfrak{so}$ $\mathfrak{weitl\ddot{a}uftig}$ $\mathfrak{ausnimmt}$, $\mathfrak{entscheidet}$ $\mathfrak{sich}$ $\mathfrak{im}$ $\mathfrak{konkreten}$ $\mathfrak{Fall}$ $\mathfrak{oft}$ $\mathfrak{auf}$ $\mathfrak{den}$ $\mathfrak{ersten}$ $\mathfrak{Blick}$: $\mathfrak{aber}$ $\mathfrak{es}$ $\mathfrak{ist}$ $\mathfrak{doch}$ $\mathfrak{der}$ $\mathfrak{Takt}$ $\mathfrak{eines}$ $\mathfrak{ge\ddot{u}bten}$ $\mathfrak{Urteils}$ $\mathfrak{dazu}$ $\mathfrak{n\ddot{o}tig}$, $\mathfrak{und}$ $\mathfrak{man}$ $\mathfrak{mu\ss}$ $\mathfrak{an}$ $\mathfrak{alle}$ $\mathfrak{die}$ $\mathfrak{hier}$ $\mathfrak{entwickelten}$ $\mathfrak{F\ddot{a}lle}$ $\mathfrak{einmal}$ $\mathfrak{gedacht}$ $\mathfrak{haben}$.\footnote{~
``\emph{What here seems so prolix in the explanation is often decided in the concrete case at first sight; but still, the tact of a practised judgment is required for that, and person must have thought over every one of the cases now developed.}'' (Source: Carl von Clausewitz: \emph{On War}, Book 6, Chapter 24 [Operating Against a Flank]. Translated by Colonel James John Graham, London, N. Tr\"ubner, 1873. \url{http://clausewitz.com/readings/OnWar1873/TOC.htm})}
}
\vspace{0.25cm}

\flushright
\noindent (Carl von Clausewitz: \emph{Vom Kriege})

\vspace{1cm} 
\justify }

\begin{document}

\newgeometry{top=20mm,bottom=25mm,left=25mm,right=25mm}
\maketitle
\thispagestyle{empty}
\Dedication

\begin{abstract}
\noindent
The outcome of some football matches has benefited both teams at the expense of a third team because head-to-head results were used for breaking ties. Inspired by these examples, our mathematical analysis identifies all possible collusion opportunities caused by this particular ranking rule in a single round-robin tournament with four teams. According to a simulation model based on the 2024 UEFA European Football Championship, merely the tie-breaking rule increases the probability of reaching a situation vulnerable to collusion by between 11.5 and 14.8 percentage points. This risk can scarcely be mitigated by a static match schedule. Therefore, tournament organisers are strongly encouraged to choose goal difference as the primary tie-breaking rule, similar to the official policy of FIFA.




\AMS{62F07, 90-10, 90B90}

\JEL{C44, C63, Z20}
\end{abstract}

\clearpage
\restoregeometry

\section{Introduction} \label{Sec1}

The expansion of the FIFA World Cup to 48 teams from 2026 has called the attention of both the scientific community and the media to the issue of tournament design. The reform was approved in 2017 together with the format of 16 groups of three teams each, where the top two teams from each group advance to the knockout stage \citep{FIFA2017d}. The decision has been strongly criticised due to the high risk of collusion or match-fixing (in line with the literature, we use the terms ``collusion'' and ``match-fixing'' as synonyms throughout the paper), and several fairer tournament formats have been proposed \citep{GuajardoKrumer2023, Guyon2020a, Palacios-Huerta2018, Renno-Costa2023}.
Finally, the organiser F\'ed\'eration Internationale de Football Association (FIFA) has decided to split the teams into 12 groups of four, with the top two from each group and the eight best third-placed teams qualifying for the Round of 32 \citep{FIFA2023b}.

However, this format does not fully eliminate the risk of match-fixing: teams may be tempted to collude in the last round if there exists a result allowing both of them to advance. If the final result lies in this appropriate ``collusion zone'', suspicions will immediately arise even if the two opposing teams have exerted full effort to score. Furthermore, other matches played in the last round may become irrelevant, which could substantially decrease their attractiveness \citep{BuraimoForrestMcHaleTena2022}. Therefore, preventing this type of match-fixing is important for financial reasons, too.

Similar tacit coordination is abundant in the history of football \citep{Guyon2020a}. The most famous example is the ``disgrace of Gij\'on'' \citep[Section~3.9.1]{KendallLenten2017}, a match played by West Germany and Austria in the 1982 FIFA World Cup, where the two teams refused to attack each other after West Germany scored a goal in the first ten minutes.

Nonetheless, it remains possible to reduce the probability of such unwanted scenarios and increase the stakes of the games played in the last round by choosing an appropriate schedule for the group matches \citep{ChaterArrondelGayantLaslier2021, Guyon2020a, Stronka2020}. Indeed, \citet[Section~5.2]{ChaterArrondelGayantLaslier2021} have already determined the optimal order of the matches in the design chosen for the 2026 FIFA World Cup.

The UEFA European Championship, contested by 24 teams since 2016, is organised in essentially the same format as the 2026 FIFA World Cup. Hence, one might think that the issue of collusion is adequately investigated here, too. But the Union of European Football Associations (UEFA) prefers head-to-head results over goal difference for tie-breaking purposes---and tie-breaking rules \emph{do} make a huge difference in single round-robin tournaments with four teams \citep{Berker2014, Csato2023a}. According to Section~\ref{Sec2}, there are some matches in the history of European football where favouring head-to-head results has created collusion opportunities.

This research gap has inspired our research questions:
(a) What are the situations where a collusion opportunity is caused \emph{only} by tie-breaking based on head-to-head results, which is the official policy of UEFA?
(b) What is the probability of these events?
(c) What is the optimal match schedule to avoid match-fixing if head-to-head results are preferred to goal difference?

Crucially, if goal difference is the primary tie-breaking rule and it implies the existence of a match-fixing situation, then the same match-fixing situation occurs if tie-breaking is based on head-to-head results. The reason is that using head-to-head results narrows the set of positions available for a team: the goal differences of any team can change arbitrarily until the end of the last round, but head-to-head results may already be fixed by the results in the previous rounds. Consequently, preferring goal difference is always optimal with respect to collusion opportunities, which is fully in line with the message of the literature on tie-breaking rules \citep{Berker2014, Csato2023a, Pakaslahti2019}. In this sense, we aim to explore the ``price'' of the UEFA tie-breaking policy in terms of match-fixing.

Therefore, our main contribution to the extant literature is the connection of two lines of research; while both tie-breaking rules \citep{Berker2014, Csato2023a} and the probability of collusion \citep{ChaterArrondelGayantLaslier2021, Guyon2020a, Stronka2020, Stronka2024} have been thoroughly investigated, their interaction has not been studied before.
The main findings can be summarised as follows:
\begin{itemize}
\item
Our mathematical model identifies all match-fixing opportunities caused by using head-to-head results to break ties in the last round of a single round-robin contest with four teams (Section~\ref{Sec4}).
\item
According to the simulation analysis of the 2024 UEFA European Championship, the probability of a match with a collusion opportunity increases substantially, by between 11.5 and 14.8 percentage points, if head-to-head results are favoured over goal difference (Section~\ref{Sec6}).
\item
In the 2024 UEFA European Championship, the proposed model uncovers that the optimal schedule in the last round of the group stage contains matches between the two strongest and the two weakest teams. However, this fixed schedule can scarcely mitigate the risk of collusion compared to a random order of the group matches (Section~\ref{Sec6}).
\end{itemize}
To conclude, the official tie-breaking policy of UEFA creates a serious fairness issue by strongly aggravating the risk of collusion in groups of four teams. Therefore, UEFA and other tournament organisers are strongly encouraged to follow the rules of FIFA and choose goal difference as the primary tie-breaker.



\section{Motivating examples} \label{Sec2}

In this section, two historical matches are revoked to highlight the serious consequences of preferring head-to-head results in tie-breaking for match-fixing opportunities.

\begin{table}[t!]
\centering
\caption{2004 UEFA European Football Championship, Group C}
\label{Table1}

\begin{subtable}{\linewidth}
\centering
\caption{Match results}
\label{Table1a}
\rowcolors{1}{}{gray!20}
    \begin{tabularx}{0.9\linewidth}{lLLc} \toprule
    Date  & First team & Second team & Result \\ \bottomrule \showrowcolors
    14 June 2014, 17:00 & Denmark & Italy & 0-0 \\
    14 June 2014, 19:45 & Sweden & Bulgaria & 5-0 \\ \hline
    18 June 2014, 17:00 & Bulgaria & Denmark & 0-2 \\
    18 June 2014, 19:45 & Italy & Sweden & 1-1 \\ \hline
    22 June 2014, 19:45 & Italy & Bulgaria & To be played \\
    22 June 2014, 19:45 & Denmark & Sweden & To be played \\ \bottomrule
    \end{tabularx}
\end{subtable}

\vspace{0.5cm}
\begin{subtable}{\linewidth}
\caption{Standing before the last matchday}
\label{Table1b}
\begin{threeparttable}
\rowcolors{3}{}{gray!20}
    \begin{tabularx}{\linewidth}{Cl CCC CCC >{\bfseries}C} \toprule \hiderowcolors
    Pos   & Team  & W     & D     & L     & GF    & GA    & GD    & Pts \\ \bottomrule \showrowcolors
    1     & Sweden & 1     & 1     & 0     & 6     & 1     & $+5$ & 4 \\
    2     & Denmark & 1     & 1     & 0     & 2     & 0     & $+2$ & 4 \\
    3     & Italy & 0     & 2     & 0     & 1     & 1     & $0$ & 2 \\
    4     & Bulgaria & 0     & 0     & 2     & 0     & 7     & $-7$ & 0 \\ \bottomrule    
    \end{tabularx}
    
    \begin{tablenotes} \footnotesize
\item
The top two teams qualify for the quarterfinals.
\item
Pos = Position; W = Won; D = Drawn; L = Lost; GF = Goals for; GA = Goals against; GD = Goal difference; Pts = Points. All teams have played two matches. 
    \end{tablenotes}
\end{threeparttable}
\end{subtable}

\end{table}

\begin{example} \citep[Section~3.9.3]{KendallLenten2017} \label{Examp1}
Table~\ref{Table1} presents Group C in the 2004 UEFA European Football Championship after two matchdays.
The top two teams advanced to the quarterfinals.
If two or more teams in the group were equal on points on completion, the following tie-breaking criteria were applied in the order given:
\begin{enumerate}
\item
Higher number of points obtained in the matches played among the teams in question;
\item
Superior goal difference in the matches played among the teams in question;
\item
Number of goals scored in the matches played among the teams in question;
\end{enumerate}
Further tie-breaking rules are not important for our discussion.

Bulgaria could have finished with at most three points. Italy would have five points by defeating Bulgaria, thus, Italy might have qualified if Denmark and Sweden did not play a draw. If Denmark and Sweden would have drawn with a score of at least 2-2, then both of them would have qualified independently of the result of Italy vs.\ Bulgaria because the head-to-head results would have favoured both Denmark and Sweden over Italy (this condition would not be satisfied by a draw of 0-0 or 1-1).
Therefore, the two Scandinavian teams had an opportunity to collude due to using head-to-head results as the primary tie-breaking principle. However, if the order of teams with equal points would have been primarily decided by goal difference among all teams including Bulgaria, such a collusion would be impossible.
\end{example}

To summarise, Denmark and Sweden knew that they would progress with a draw of 2-2. The final result of the match was 2-2, although both teams denied any accusation.


\begin{table}[t!]
\centering
\caption{2017 UEFA European Under-21 Championship, Group C}
\label{Table2}

\begin{subtable}{\linewidth}
\centering
\caption{Match results}
\label{Table2a}
\rowcolors{1}{}{gray!20}
    \begin{tabularx}{0.9\linewidth}{lLLc} \toprule
    Date  & First team & Second team & Result \\ \bottomrule \showrowcolors
    18 June 2017, 18:00 & Germany & Czech Republic & 2-0 \\
    18 June 2017, 20:45 & Denmark & Italy & 0-2 \\ \hline
    21 June 2017, 18:00 & Czech Republic & Italy & 3-1 \\
    21 June 2017, 20:45 & Germany & Denmark & 3-0 \\ \hline
    24 June 2014, 19:45 & Italy & Germany & To be played \\
    24 June 2014, 19:45 & Czech Republic & Denmark & To be played \\ \bottomrule
    \end{tabularx}
\end{subtable}

\vspace{0.5cm}
\begin{subtable}{\linewidth}
\caption{Standing before the last matchday}
\label{Table2b}
\begin{threeparttable}
\rowcolors{3}{}{gray!20}
    \begin{tabularx}{\linewidth}{Cl CCC CCC >{\bfseries}C} \toprule \hiderowcolors
    Pos   & Team  & W     & D     & L     & GF    & GA    & GD    & Pts \\ \bottomrule \showrowcolors
    1     & Germany & 2     & 0     & 0     & 5     & 0     & $+5$ & 6 \\
    2     & Czech Republic & 1     & 0     & 1     & 3     & 3     & $0$ & 3 \\
    3     & Italy & 1     & 0     & 1     & 3     & 3     & $0$ & 3 \\
    4     & Denmark & 0     & 0     & 2     & 0     & 5     & $-5$ & 0 \\ \bottomrule    
    \end{tabularx}
    
    \begin{tablenotes} \footnotesize
\item
The group winner qualifies for the semifinals.
The best runner-up of groups A--C also qualifies for the semifinals. The best runner-up from Groups A and B is Slovakia with 6 goals for, 3 goals against, and a goal difference of $+3$.
\item
Pos = Position; W = Won; D = Drawn; L = Lost; GF = Goals for; GA = Goals against; GD = Goal difference; Pts = Points. All teams have played two matches.    
    \end{tablenotes}
\end{threeparttable}
\end{subtable}

\end{table}

\begin{example} \label{Examp2}
Table~\ref{Table2} shows Group C in the 2017 UEFA European Under-21 Championship after two matchdays.
The primary tie-breaking criteria were analogous to the 2004 UEFA European Football Championship, see Example~\ref{Examp1}.

Besides the three group winners, the best runner-up also advanced to the semifinals. Obviously, the runners-up were evaluated based on their number of points, followed by goal difference. The other two groups were already finished, and the better second-placed team was Slovakia, which scored 6 and conceded 3 goals, implying a goal difference of $+3$.

In this situation, Italy and Germany had an opportunity to collude. If Italy defeats Germany by 4-2, then two cases should be distinguished:
\begin{enumerate}[label = (\arabic*)]
\item
If the Czech Republic wins against Denmark, three teams will have 6 points. The head-to-head goal differences will be 3-3 (Czech Republic), 5-5 (Italy), 4-4 (Germany). Therefore, Italy will qualify as the group winner and Germany as the runner-up due to scoring more goals than Slovakia.
\item
If the Czech Republic does not win against Denmark, Italy will qualify as the group winner and Germany as the runner-up.
\end{enumerate}
\end{example}

Finally, Italy beat Germany 1-0 and Denmark defeated the Czech Republic. Therefore, Italy and Germany progressed at the expense of Slovakia since Italy was the group winner and Germany was the best runner-up ahead of Slovakia as Germany had a goal difference of $+4$. However, Slovakia had the same number of points but a superior goal difference ($+3$) compared to the group winner Italy ($+1$).
The prime minister of Slovakia wrote an open letter to UEFA, emphasising that ``\emph{I believe that you will have the whole issue investigated and UEFA will make rules for the next tournament which will give precedence to a sportsmanlike performance instead of unfair agreements}'' \citep{BBC2017}.

According to Examples~\ref{Examp1} and \ref{Examp2}, preferring head-to-head results to goal difference may create additional match-fixing opportunities: in both cases, the potential collusion would have been impossible if ties had been broken by goal difference. This aspect has not been considered in the previous literature but will be addressed here.

\section{Related literature} \label{Sec3}

\citet{Guyon2020a} examines the risk of match-fixing in a single round-robin tournament with three teams, the format originally suggested for the 2026 FIFA World Cup.
In a round-robin tournament with four teams, \citet{Stronka2020} uncovers the role of the schedule with respect to the temptation to deliberately lose, which arises from the structure of the subsequent knockout stage.
\citet{ChaterArrondelGayantLaslier2021} develop a classification method to quantify the attractiveness of the games played in the last round of a round-robin competition. The authors calibrate various score-predicting models based on historical data to determine the schedule that minimises the occurrence of match-fixing opportunities in single round-robin tournaments with three, four, and five teams.
\citet{Stronka2024} demonstrates that a substantial reduction in collusion risk can be achieved in the case of three teams by random tie-breaking (a biased lottery based on goal differences) and dynamic scheduling (when the order of the games is not fixed in advance but depends on the results of previous games).

These papers already highlight the importance of the match schedule for match-fixing opportunities.
On the other hand, they assume that all ties are broken by goal difference instead of head-to-head results. This is not a problem in single round-robin tournaments with three teams, where the two choices are equivalent. But the tie-breaking rule can have fundamental effects if there are four teams, and the previous literature completely ignores the situations described in Examples~\ref{Examp1} and \ref{Examp2}.

We also contribute to the relatively limited literature on tie-breaking rules for sports tournaments. \citet{Berker2014} calculates how often the relative ranking of two teams depends on the outcome of a game played by other teams under the two main tie-breaking concepts. According to the philosophical arguments of \citet{Pakaslahti2019}, tie-breaking systems of round-robin football competitions should give more importance to overall goal difference than to head-to-head results. \citet[Chapter~1.3]{Csato2021a} reveals the lack of consensus around tie-breaking criteria in the top tier football leagues in Europe. \citet{Csato2023a} compares goal difference and head-to-head results with respect to the probability that the final rank of a team becomes known before the last round.

The effect of the schedule on the outcome of a tournament is investigated by \citet{KrumerMegidishSela2017a}. It is shown that, in round-robin contests with three or four symmetric contestants, there is a first-mover advantage driven by strategic effects arising from the subgame perfect equilibrium. The theoretical finding of first-mover advantage holds for real-world tournaments, including the FIFA World Cup \citep{KrumerLechner2017}.

Match-fixing arising from the qualification of multiple teams can be partially resolved by changing the pairing rule in the subsequent knockout stage of the tournament. \citet{Stronka2020} suggests the unanimity method, where group winners are allowed to play against each other rather than the runner-up. According to the proposal of \citet{Guyon2022a}, teams that have performed best during the group stage can sequentially choose their opponents subject to some constraints. Analogously, the solution of \citet{HallLiu2024} allows higher-ranked players to choose their next opponent at each round of the single elimination phase.

Our mathematical model takes an \emph{ex ante} perspective, that is, it focuses on games that may be vulnerable to match-fixing. A recent line of research examines the same problem \emph{ex post} by looking for statistical evidence of match-fixing \citep{ForrestMcHale2019, ScellesFrancoisValenti2024}.

Last but not least, if the preference for head-to-head results creates a collusion opportunity, all other matches become irrelevant. Hence, our study is connected to the substantial literature on match importance in round-robin tournaments \citep{BuraimoForrestMcHaleTena2022, FaellaSauro2021, Geenens2014, GollerHeininger2023, GoossensBelienSpieksma2012, ScarfShi2008}.

\section{Theoretical analysis} \label{Sec4}

This section will identify all match-fixing opportunities caused by using head-to-head results for tie-breaking in a single round-robin tournament with four teams. For instance, the case when two teams have 2-2 wins after two games such that they play against each other in the final round is \emph{not} studied because the existence of collusion by playing a draw does not depend on the tie-breaking rule.

The two matches in the last round are assumed to be played simultaneously, which is a standard rule in major football tournaments since the ``disgrace of Gij\'on'' \citep[Section~3.9.1]{KendallLenten2017}. A win is awarded by three, a draw by one, and a loss by zero points.

Clearly, tacit collusion is not possible in the first two rounds because of the substantial uncertainty due to the unknown results of the remaining matches.
Two teams $A$ and $B$ can exploit the use of head-to-head results as primary tie-breaking criteria to collude in the final round if:
(a) exactly three teams, including $A$ and $B$, can be equal on points on completion; and
(b) they play against each other in the last round.
Note that in a single group setting, the tie-breaking rule cannot be detrimental to a third team if only teams $A$ and $B$ have the same number of points. Analogously, if four teams have the same number of points, the two tie-breaking concepts are equivalent.

Both of these teams should finish with 2, 3, 4, 5, or 6 points. Then, all match-fixing opportunities can be derived by ``backward induction''.

\begin{enumerate}
\item \label{Case1}
Teams $A$ and $B$ finish with 6 points \\
Before the last round, $A$ has had 6 and $B$ has had 3 points such that the goal difference of the other team $C$ with 3 points has not been positive. \\
The scenario is called \emph{Robust collusion 1} (RC1) because the game $A$ vs.\ $B$ has an outcome (team $B$ wins with an appropriate number of goals) which ensures that teams $A$ and $B$ will be the top two teams, independently of the result of the other match played in the last round. 
\begin{proof}
Let the results of the matches $A$ vs.\ $B$, $A$ vs.\ $C$, $B$ vs.\ $C$ be $s$-$t$, $u$-$v$, $x$-$y$, respectively, where $s > t$, $u > v$, $x < y$. If $C$ does not win its last match, $A$ and $B$ will be the top two teams. If $C$ wins its last match, then $A$, $B$, $C$ will have 6 points each, and head-to-head goal differences will break the tie. These are $s+u-t-v$ for $A$, $t+x-s-y$ for $B$, $v+y-u-x$ for $C$, respectively. Since the goal difference of team $C$ is not positive ($v+y-u-x \leq 0$), there exist integers $s,t \geq 0$ such that $s+u-t-v \geq 0$ and $t+x-s-y \geq 0$, furthermore, $s+u > v+y$ and $t+x > v+y$, which guarantee that both $A$ and $B$ are ranked above $C$.
\end{proof}
This situation is analogous to Example~\ref{Examp2} with team $A$ being Germany, team $B$ being Italy, and team $C$ being the Czech Republic with a goal difference of $0$, see Table~\ref{Table2b}.

\item \label{Case2}
Teams $A$ and $B$ finish with 5 points \\
Before the last round, both $A$ and $B$ have had 4 points, and a third team has had 2 points. \\
The scenario is called \emph{Robust collusion 2} (RC2) because the game $A$ vs.\ $B$ has an outcome (a draw with a sufficiently high number of goals) which ensures that teams $A$ and $B$ will be the top two teams, independently of the result of the other match played in the last round.
\begin{proof}
Let the results of the matches $A$ vs.\ $B$, $A$ vs.\ $C$, $B$ vs.\ $C$ be $s$-$s$, $t$-$t$, $u$-$u$, respectively. If $C$ does not win its last match, $A$ and $B$ will be the top two teams. If $C$ wins its last match, then $A$, $B$, $C$ will have 5 points each, and number of head-to-head goals scored will break the tie. These are $s+t$ for $A$, $s+u$ for $B$, $t+u$ for $C$, respectively. There exists an integer $s \geq 0$ such that $s+t > t+u$ and $s+u > t+u$ (e.g.\ $s = \max \{ t;u \} +1$), which guarantee that both $A$ and $B$ are ranked higher than $C$.
\end{proof}
This situation is analogous to Example~\ref{Examp1} with teams $A$, $B$, $C$ being Denmark, Sweden, and Italy, respectively, see Table~\ref{Table1b}.

Note that, if $A$ and $B$ have had 4 points after two rounds, the other two teams might have 1 point each, too. But then teams $A$ and $B$ can secure the first two positions for them by playing a draw, independently of the tie-breaking criteria.

\item \label{Case3}
Teams $A$ and $B$ finish with 4 points \\
Before the last round, $A$ has had 4 points and $B$ has had 1 point such that the goal difference of the team with 3 points has not been positive. \\
The scenario is called \emph{Conditional collusion 1} (CC1) because the game $A$ vs.\ $B$ has an outcome (team $B$ wins with an appropriate number of goals), which ensures that teams $A$ and $B$ will be the top two teams if the other match played in the last round is a draw.
\begin{proof}
Analogous to the proof of RC1.
\end{proof}

\item \label{Case4}
Teams $A$ and $B$ finish with 3 points \\
Before the last round, $A$ has had 3 and $B$ has had 0 points such that the goal difference of the other team with 3 points has not been positive. \\
The scenario is called \emph{Conditional collusion 2} (CC2) because the game $A$ vs.\ $B$ has an outcome (team $B$ wins with an appropriate number of goals), which ensures that teams $A$ and $B$ will be the runner-up and the third-placed team if the other match played in the last round is won by the team which has had 6 points after two rounds due to defeating both $A$ and $B$.
\begin{proof}
Analogous to the proof of RC1.
\end{proof}
This situation emerged in Group B in the 2020 UEFA European Football Championship. In the first two rounds, Belgium defeated both Finland and Russia, while Finland vs.\ Denmark was 1-0 and Russia vs.\ Finland was 1-0. Therefore, the conditions above hold with Russia being team $A$ and Denmark being team $B$ since the goal difference of Finland was $0$. Denmark and Russia had a collusion strategy---for example, a Danish win by 2-1---to guarantee the second and third positions for these teams if Finland had lost to Belgium as happened.

\item \label{Case5}
Teams $A$ and $B$ finish with 2 points \\
Before the last round, both $A$ and $B$ have had 1 point. \\
The scenario is called \emph{Conditional collusion 3} (CC3) because the game $A$ vs.\ $B$ has an outcome (a draw with a high number of goals), which ensures that teams $A$ and $B$ will be the runner-up and the third-placed team if the other match played in the last round is won by the team which has had 6 points after two rounds due to defeating both $A$ and $B$.
\begin{proof}
Analogous to the proof of RC2.
\end{proof}
\end{enumerate}
Finally, the conditions of RC1 and CC2 might hold simultaneously if the team having 6 points plays against a team having 3 points in the last round (consequently, the other team having 3 points plays against the team having 0 points) and the goal differences of both teams with 3 points are not positive. This scenario is called \emph{Dual collusion} (DC).

\begin{table}[t!]
  \centering
  \caption{Conditions for different types of match-fixing before the last round}
  \label{Table3}
  \rowcolors{3}{}{gray!20} 
\begin{threeparttable}
    \begin{tabularx}{\textwidth}{L CCcc} \toprule
    Type of collusion & Point distribution & Draws & Games to be played & Goal difference \\ \bottomrule \showrowcolors
    RC1   & [6, 3, 3, 0] & 0 & (6 vs 3); (3 vs 0) & $\leq 0$ for 3 in (3 vs 0) \\
    RC2   & [4, 4, 2, 0] & 2 & (4 vs 4); (2 vs 0) & --- \\
    CC1   & [4, 3, 2, 1] & 2 & (4 vs 1); (3 vs 2) & $\leq 0$ for 3 \\
    CC2   & [6, 3, 3, 0] & 0 & (6 vs 3); (3 vs 0) & $\leq 0$ for 3 in (6 vs 3) \\
    CC3   & [6, 2, 1, 1] & 2 & (6 vs 2); (1 vs 1) & --- \\ \hiderowcolors
    \multirow{2}{*}{DC}    & \multirow{2}{*}{[6, 3, 3, 0]} & \multirow{2}{*}{0} & \multirow{2}{*}{(6 vs 3); (3 vs 0)} & $\leq 0$ for 3 in (3 vs 0) \\
        & & & & $\leq 0$ for 3 in (6 vs 3) \\ \bottomrule
    \end{tabularx}
\begin{tablenotes} \footnotesize
\item
The column Draws shows the number of draws in the first two rounds of the tournament.
\item
In the last two columns, the teams are denoted by their number of points scored.
\item
Tacit collusion ensures the top two positions in RC1, RC2, CC1, and the second and third positions in CC2, CC3.
\item
DC stands for simultaneous RC1 and CC2.
\item
Goal difference is irrelevant for types RC2 and CC3 because head-to-head results serve as the primary tie-breaking criteria.
\end{tablenotes}
\end{threeparttable}
\end{table}

All types of match-fixing opportunities are summarised in Table~\ref{Table3}.
Although conditional collusion is less serious than robust collusion, since the two matches in the last round are usually played simultaneously and the actual result of the other match is immediately available today, we think these situations are also relevant for tournament organisers.

The same calculations can be followed to identify match-fixing opportunities created by using head-to-head results in the final round of round-robin tournaments containing a higher number of rounds or teams. However, the point distribution before the last round becomes more complex, and collusion may arise in previous rounds.
Note also that, as it has been mentioned in Section~\ref{Sec3}, the two widely used tie-breaking principles (head-to-head results and goal difference) coincide in groups of three.

\section{The simulation model} \label{Sec5}

In order to quantify the seriousness of collusion opportunities derived in Section~\ref{Sec4}, we will use Monte Carlo simulations. Our case study is provided by the 2024 UEFA European Football Championship (shortly, the UEFA Euro 2024), where the group winners, the runners-up, and two-thirds of the third-placed teams qualify from the six round-robin groups of four teams, and the primary tie-breaking rule is head-to-head results \citep[Article~15]{UEFA2022g}---similar to Example~\ref{Examp1}. Consequently, all types of collusion opportunities identified in Section~\ref{Sec4} might be relevant.

\begin{table}[t!]
  \centering
  \caption{The groups of the 2024 UEFA European Football Championship}
  \label{Table4} 
\rowcolors{3}{gray!20}{} 
    \begin{tabularx}{\textwidth}{Lc m{0.5cm} Lc m{0.5cm} Lc} \toprule \hiderowcolors
    \multicolumn{2}{c}{\textbf{Group A}} &       & \multicolumn{2}{c}{\textbf{Group B}} &       & \multicolumn{2}{c}{\textbf{Group C}} \\ 
    Team 	& Elo   &       & Team & Elo &       & Team & Elo \\ \bottomrule \showrowcolors
    Germany & 1886  &       & Spain & 2033  &       & England & 2015 \\
    Hungary & 1834  &       & Albania & 1632  &       & Denmark & 1825 \\
    Scotland & 1801  &       & Croatia & 1952  &       & Slovenia & 1710 \\
    Switzerland & 1792  &       & Italy & 1938  &       & Serbia & 1802 \\ \bottomrule
	\end{tabularx}

\vspace{0.25cm}
\begin{threeparttable}
\rowcolors{3}{gray!20}{}	
    \begin{tabularx}{\textwidth}{Lc m{0.5cm} Lc m{0.5cm} Lc} \toprule \hiderowcolors
    \multicolumn{2}{c}{\textbf{Group D}} &       & \multicolumn{2}{c}{\textbf{Group E}} &       & \multicolumn{2}{c}{\textbf{Group F}} \\ 
    Team 	& Elo   &       & Team & Elo &       & Team & Elo \\ \bottomrule \showrowcolors
    France & 2110  &       & Belgium & 1990  &       & Portugal & 2033 \\
    Austria & 1835  &       & Romania & 1674  &       & Turkey & 1766 \\
    Netherlands & 1970  &       & Slovakia & 1662  &       & Czech Republic & 1757 \\
    Play-off A & 1710  &       & Play-off B & 1850  &       & Play-off C & 1729 \\ \bottomrule
    \end{tabularx}
\begin{tablenotes} \footnotesize
\item
Groups D, E, F contain the winners of the play-off paths A, B, C, respectively, that are unknown at the time of the draw. The winners are assumed to be the strongest teams in the play-off paths: Poland (Play-off Path A), Ukraine (Play-off Path B), (Play-off Path C).
\item
The strengths of the teams are measured by their World Football Elo Ratings on 1 December 2023 (the date of the group draw), see \url{https://www.international-football.net/elo-ratings-table?year=2023&month=12&day=01&confed=UEFA}.
\end{tablenotes}
\end{threeparttable}
\end{table}

The compositions of the groups are shown in Table~\ref{Table4}. The six groups provide a good mixture of teams: Group B has one strong, two middle, and one weak team, Groups A, C and F have one strong, and three weak teams, while Groups D and E have one strong, one middle, and one weak team.

Tournament designs are usually evaluated by simulating the matches on the basis of a reasonable statistical model \citep{ScarfYusofBilbao2009}. Since the number of goals scored and conceded is of high importance to us, the traditional Poisson model is used to obtain these values \citep{HubacekSourekZelezny2022, LeyVandeWieleVanEeetvelde2019, Maher1982, ScarfKhareAlotaibi2022, VanEetveldeLey2019}.
If a match is played on field $f$ (home, neutral, or away), team $i$ scores $k$ goals against team $j$ with a probability of
\begin{equation} \label{Poisson_dist}
P_{ij}(k) = \frac{ \left( \lambda_{ij}^{(f)} \right)^k \exp \left( -\lambda_{ij}^{(f)} \right)}{k!},
\end{equation}
where $\lambda_{ij}^{(f)}$ is the expected number of goals scored by team $i$ in this particular match.

The strength of a team is measured by its World Football Elo Rating as it is widely accepted in the literature \citep{CeaDuranGuajardoSureSiebertZamorano2020, GasquezRoyuela2016, HvattumArntzen2010, LasekSzlavikBhulai2013}. The Elo ratings of the teams on the day of the UEFA Euro 2024 draw are shown in Table~\ref{Table4}. The win expectancy of team $i$ against team $j$ is
\[
W_{ij} = \frac{1}{1 + 10^{-(E_i - E_j)/400}},
\]
where $E_i$ and $E_j$ are the Elo ratings of teams $i$ and $j$, respectively, such that the rating of the home team (in our case, Germany) is increased by 100 points.

\citet{FootballRankings2020} has estimated the parameter $\lambda_{ij}^{(f)}$ in \eqref{Poisson_dist} as the function of win expectancy $W_{ij}$ based on more than 40 thousand matches played by national football teams.
The expected number of goals for the home team $i$ equals
\begin{equation} \label{Exp_goals_home}
\lambda_{ij}^{(h)} = 
\left\{ \begin{array}{ll}
-5.42301 \cdot W_{ij}^4 + 15.49728 \cdot W_{ij}^3 \\
- 12.6499 \cdot W_{ij}^2 + 5.36198 \cdot W_{ij} + 0.22862 & \textrm{if } W_{ij} \leq 0.9 \\ \\
231098.16153 \cdot (W_{ij}-0.9)^4 - 30953.10199 \cdot (W_{ij}-0.9)^3 & \\
+ 1347.51495 \cdot (W_{ij}-0.9)^2 - 1.63074 \cdot (W_{ij}-0.9) + 2.54747 & \textrm{if } W_{ij} > 0.9,
\end{array} \right.
\end{equation}
and the expected number of goals for the away team $j$ is
\begin{equation} \label{Exp_goals_away}
\lambda_{ij}^{(a)} = 
\left\{ \begin{array}{ll}
90173.57949 \cdot (W_{ij} - 0.1)^4 + 10064.38612 \cdot (W_{ij} - 0.1)^3 \\
+ 218.6628 \cdot (W_{ij} - 0.1)^2 - 11.06198 \cdot (W_{ij} - 0.1) + 2.28291 & \textrm{if } W_{ij} < 0.1 \\ \\
-1.25010 \cdot W_{ij}^4 -  1.99984 \cdot W_{ij}^3 & \\
+ 6.54946 \cdot W_{ij}^2 - 5.83979 \cdot W_{ij} + 2.80352 & \textrm{if } W_{ij} \geq 0.1.
\end{array} \right.
\end{equation}
However, these formulas are valid only for the two matches played by Germany in Group A since the last round of group matches is not simulated.

Most games are played on a neutral field when the expected number of goals for team $i$ against team $j$ is
\begin{equation} \label{Exp_goals_neutral}
\lambda_{ij}^{(n)} = 
\left\{ \begin{array}{ll}
3.90388 \cdot W_{ij}^4 - 0.58486 \cdot W_{ij}^3 \\
- 2.98315 \cdot W_{ij}^2 + 3.13160 \cdot W_{ij} + 0.33193 & \textrm{if } W_{ij} \leq 0.9 \\ \\
308097.45501 \cdot (W_{ij}-0.9)^4 - 42803.04696 \cdot (W_{ij}-0.9)^3 & \\
+ 2116.35304 \cdot (W_{ij}-0.9)^2 - 9.61869 \cdot (W_{ij}-0.9) + 2.86899 & \textrm{if } W_{ij} > 0.9,
\end{array} \right.
\end{equation}
with $R^2 = 0.976$.

\citet{Csato2022a, Csato2023d} and \citet{Stronka2024} have used the same approach recently.

In all groups, the outcomes of the games are simulated according to the model above in order to obtain the probability of each type of collusion opportunity. For robust collusion, it is sufficient to simulate the first two rounds. However, in the case of conditional collusion, we take the result of the other match played in the last round into account and consider the opportunity valid only if the outcome of that match is the required one (for example, it is a draw if the conditions of CC3 hold).

The number of iterations $N$ affects the reliability of the results. Since the probability $p$ of an event is estimated by its relative frequency in the simulation, the margin of error required for the 99\% confidence interval is approximately $2.8 \sqrt{p(1-p)} / \sqrt{N}$. Due to our choice of $N=1$ million, it is sufficient to focus on relative frequencies because the margin of error is at most $\left( 2.8 \times \sqrt{0.5 \times 0.5} \right) / 1000 = 0.007$\%.

The organiser can decide the order of group matches. Denote the four teams by $T1$--$T4$ according to their strengths. Each team plays three games in the group, thus, three scheduling options exist for the last round, where potential match-fixing may arise:
\begin{itemize}
\item
Schedule S1: $T1$ plays against $T2$ and $T3$ plays against $T4$;
\item
Schedule S2: $T1$ plays against $T3$ and $T2$ plays against $T4$;
\item
Schedule S3: $T1$ plays against $T4$ and $T2$ plays against $T3$.
\end{itemize}
Table~\ref{Table4} uncovers that the Elo rating sometimes implies a quite different ranking compared to the official seeding. In these cases, the teams are labelled according to their Elo ratings since our simulation model does use them as a measure of strength.

The simulations have been implemented in Python 3.7.7. 
The computation time of the 1 million simulation runs is below 30 minutes on a standard laptop for all the six groups and one given schedule. Computation time increases linearly with the number of iterations and the number of schedules but it does not make much sense to choose more than 1 million runs as discussed above.

\section{The relative frequency of collusion opportunities} \label{Sec6}

\input{Figure1_collusion_schedule}

Figure~\ref{Fig1} presents the main results, the probability of different types of match-fixing caused by preferring head-to-head results under the three possible schedules in each single round-robin group of the UEFA Euro 2024.
They can be summarised as follows:
\begin{itemize}
\item
Robust collusion 1 has the highest probability, at least 10\%, which exceeds even the worst case for all other types of collusion;
\item
The threat of Robust collusion 2 cannot be neglected as its chance remains above 2.5\% and can easily lead to controversies according to Example~\ref{Examp1};
\item
Conditional collusion 1 is the least likely to occur due to requiring three draws out of five matches;
\item
Conditional collusion 2 is the second most probable event with a chance over 4\% and can be especially harmful in UEFA Euro, where the second- and third-placed teams qualify for the Round of 16 in two-thirds of the groups;
\item
The probability of Conditional Collusion 3 remains rather low, and it is unlikely that both colluding teams advance to the knockout stage with two points each;
\item
With a probability of at least 2.4\%, both group matches played in the last round create incentives for match-fixing (Dual collusion).
\end{itemize}

The cumulated probability of robust collusion can be reduced by 18\% (Group F) to 44\% (Group D) if Schedule S3 is chosen instead of S1. For example, the overall probability of RC1 and RC2 is 17.9\% under Schedule S1 and 14.7\% under Schedule S3, implying a relative advantage of 18\% for S3.
However, compared to Schedule S1, Schedule S3 more than doubles the likelihood of Conditional collusion 2 in Groups D and E, which contain one strong, one middle, and two weak teams. Schedule S1 performs the best for the other three types of collusion, including Dual collusion, when both matches played in the last round may inspire collusion. Furthermore, while robust collusion seems less relevant in the current format of the UEFA Euro as 5 points almost certainly remain sufficient to qualify for the Round of 16, Conditional collusion 2 can pose a real problem for this tournament.

The optimality of a given schedule can be explained as follows. 
Under Schedule S3, the strongest and the weakest teams play against each other in the last round. Thus, it has the lowest probability that one team, probably the best, already has two wins and six points, which is a necessary condition of RC1, and RC1 has a much higher chance of occurring than RC2. Furthermore, there is essentially no difference between Schedules S2 and S3 in Group E, where the two weakest teams (Romania and Slovakia) have almost the same Elo rating.

On the other hand, Conditional collusion 2 has the lowest probability under Schedule S1 because it requires a team $t$ with three points and a negative goal difference to play against a team with six points, which is the best team in most of the cases. It is the least likely that team $t$ has one win and one more severe loss against two of the three weakest teams if it is the second best, especially if the two weakest teams have approximately the same strength as in Groups D and E.
Finally, the schedule has almost no effect on the probability of collusion in Groups A and F, which have a clear favourite (Germany and Portugal) and three other teams with nearly identical strengths.

The last round contains two group matches and the conditions of RC1, RC2, CC1, CC2, CC3 cannot hold for the same game because either the outcomes of the previous games differ, or they affect separate games if the conditions of both RC1 and CC2 are satisfied. Thus, the overall probability of a game that is vulnerable to collusion because of the UEFA tie-breaking rule---the sum of the five probabilities in Figure~\ref{Fig1} except for dual collusion DC---varies between 11.5\% and 14.8\% in the 18 scenarios (six groups with three possible schedules). The results are remarkably robust with respect to both the schedule of matches and the (relative) strengths of the teams.
This is a high value compared to the chance of collusive behaviour found for the FIFA World Cup: at most 1.8\% if the first two teams qualify \citep[Table~8]{ChaterArrondelGayantLaslier2021} and less than 9.5\% if the first three teams qualify \citep[Table~12]{ChaterArrondelGayantLaslier2021}.

In the evaluation of scheduling options for the UEFA Euro, two aspects are worth consideration:
\begin{itemize}
\item
Collusion remains possible even if the primary tie-breaking rule is goal difference. According to \citet[Table~12]{ChaterArrondelGayantLaslier2021}, the probability of collusion is 2.82\% under Schedule S1, 6.77\% under Schedule S2, and 9.50\% under Schedule S3 in the format of the 2026 FIFA World Cup, which essentially duplicates the current design of the UEFA Euro.
\item
As explained above, Conditional collusion 2 is more threatening than robust collusion.
\end{itemize}
Consequently, Schedule S1 is clearly the best choice for UEFA Euro.

To conclude, the role of tie-breaking cannot be neglected in the investigation of competitiveness in single round-robin groups with four teams.
Since preferring head-to-head results strongly aggravates the risk of collusion, goal difference should be favoured in all future tournaments, especially if the number of rounds and teams is small. This recommendation reinforces the arguments of the extant literature \citep{Berker2014, Csato2023a, Pakaslahti2019}.

\section{Conclusions} \label{Sec7}

This study has investigated how preferring head-to-head results among the tie-breaking criteria creates additional collusion and match-fixing opportunities. Situations, when two opposing teams can guarantee certain positions for them with a particular outcome of a game, need to be avoided because it is impossible to prove that the teams played competitively if the beneficial result occurs. Inspired by two real-world examples from European football, we have identified five types of potential match-fixing in the last round of a single round-robin tournament with four teams, all of them caused by using head-to-head results to break ties among teams with the same number of points.

In particular, two teams can guarantee the first two positions for them in two types of tacit collusion (RC1 and RC2) before the last round, independently of the result of the simultaneous game played by the other two teams. Three further types of match-fixing are conditional on the result of the parallel game, one of them gives the top two positions for the colluders (CC1), and the other two provide the second and third places (CC2 and CC3). These instances of match-fixing might seem redundant as usually the group winner and the runner-up qualify, but they are highly relevant for the UEFA European Championship where the four best third-placed teams advance to the Round of 16.

The six groups of the UEFA Euro 2024 have been simulated in order to estimate the probability of each collusion opportunity as the function of the game schedule. Favouring head-to-head results over goal difference substantially increases the risk of collusion in a match played in the last round, by between 11.5 and 14.8 percentage points. Even though the corresponding number for a round-robin group containing three teams is about 35\% \citep[Table~2]{Stronka2024}, our finding is only due to the misaligned tie-breaking policy of UEFA.
An appropriate schedule offers a partial remedy: to minimise the risk of collusion, the two strongest teams should play against each other in the final round.

While the results of the simulation depend on the quality of the statistical model and the distribution of teams' strengths to some extent, the study of six different groups provides a kind of sensitivity analysis, and the findings are similar for all cases. In particular, the order of the matches has a relatively low effect on collusion opportunities created by the UEFA tie-breaking rule. This may call for preferring goal difference. The identification of match-fixing opportunities in Section~\ref{Sec4} remains valid for all single round-robin contests with four teams, which is one of the most popular tournament formats in sports. Furthermore, increasing the number of matches (e.g.\ groups with six teams or double round-robin) strongly reduces the significance of the tie-breaking rule, so it is unlikely that the probability of collusion opportunities grows substantially compared to the results of \citet{ChaterArrondelGayantLaslier2021}.

The current paper reinforces a crucial message of the academic literature for tournament organisers that collusion opportunities in the final round of a round-robin contest need to be considered. Since previous research has neglected the role of tie-breaking rules, we have verified that the optimal order of the games does not change in a single round-robin tournament with four teams if three teams advance and head-to-head results are the primary tie-breaking criteria instead of goal difference. Nonetheless, favouring head-to-head results severely aggravates the risk of collusion.

Our findings can be directly used by decision-makers to increase the competitiveness of sports tournaments. Since the recent modification of the knockout bracket in the UEFA Euro has been based on a rigorous scientific study \citep{Guyon2018a}, UEFA is hopefully open to similar design changes in order to address fairness issues.

\section*{Acknowledgements}
\addcontentsline{toc}{section}{Acknowledgements}
\noindent
This paper could not have been written without \emph{my father} (also called \emph{L\'aszl\'o Csat\'o}), who has primarily coded the simulations in Python. \\
\emph{Phil Scarf} and ten colleagues and anonymous reviewers provided valuable comments and suggestions on earlier drafts. \\
We are indebted to the \href{https://en.wikipedia.org/wiki/Wikipedia_community}{Wikipedia community} for summarising important details of the sports competition discussed in the paper. \\
The research was supported by the National Research, Development and Innovation Office under Grant FK 145838.

\bibliographystyle{apalike}
\bibliography{All_references}

\end{document}

%% file: Figure1_collusion_schedule.tex
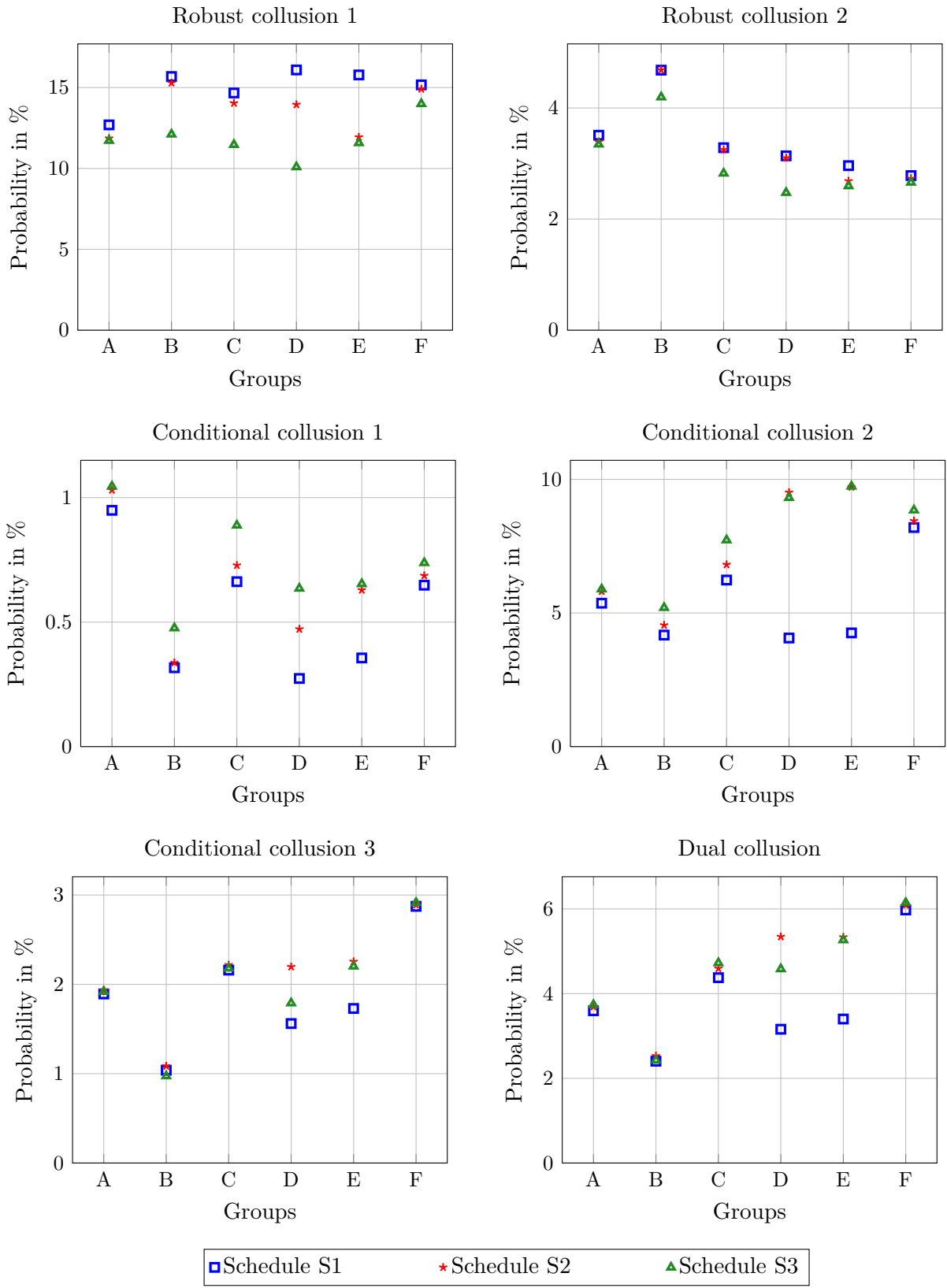
\begin{figure}[t!]
\centering

\begin{tikzpicture}
\begin{axis}[
name = axis1,
width = 0.49\textwidth, 
height = 0.4\textwidth,
title = {Robust collusion 1},
title style = {align=center, font=\small},
xmajorgrids = true,
ymajorgrids = true,
symbolic x coords = {A,B,C,D,E,F},
xtick = data,
xlabel = {Groups},
xlabel style = {align=center, font=\small},
ymin = 0,
scaled y ticks = false,
ylabel = {Probability in \%},
ylabel style = {align=center, font=\small},
yticklabel style = {/pgf/number format/fixed,/pgf/number format/precision=5},
ytick style = {draw = none},
]
\addplot [blue, only marks, mark = square, very thick] coordinates{
(A,12.6925)
(B,15.6812)
(C,14.6656)
(D,16.0886)
(E,15.78)
(F,15.1659)
};
\addplot [red, only marks, mark = star, thick] coordinates{
(A,11.8597)
(B,15.2867)
(C,14.0356)
(D,13.9456)
(E,11.9336)
(F,14.9096)
};
\addplot [ForestGreen, only marks, mark = triangle, very thick] coordinates{
(A,11.7198)
(B,12.112)
(C,11.4676)
(D,10.089)
(E,11.578)
(F,13.9924)
};
\end{axis}

\begin{axis}[
at = {(axis1.south east)},
xshift = 0.12\textwidth,
width = 0.49\textwidth, 
height = 0.4\textwidth,
title = {Robust collusion 2},
title style = {align=center, font=\small},
xmajorgrids = true,
ymajorgrids = true,
symbolic x coords = {A,B,C,D,E,F},
xtick = data,
xlabel = {Groups},
xlabel style = {align=center, font=\small},
ymin = 0,
scaled y ticks = false,
ylabel = {Probability in \%},
ylabel style = {align=center, font=\small},
yticklabel style = {/pgf/number format/fixed,/pgf/number format/precision=5},
ytick style = {draw = none},
]
\addplot [blue, only marks, mark = square, very thick] coordinates{
(A,3.5091)
(B,4.6809)
(C,3.2848)
(D,3.1384)
(E,2.962)
(F,2.7829)
};
\addplot [red, only marks, mark = star, thick] coordinates{
(A,3.3922)
(B,4.6848)
(C,3.2489)
(D,3.1021)
(E,2.6841)
(F,2.7307)
};
\addplot [ForestGreen, only marks, mark = triangle, very thick] coordinates{
(A,3.3473)
(B,4.1953)
(C,2.8234)
(D,2.4757)
(E,2.596)
(F,2.6602)
};
\end{axis}
\end{tikzpicture}

\vspace{0.25cm}
\begin{tikzpicture}
\begin{axis}[
name = axis3,
width = 0.49\textwidth, 
height = 0.4\textwidth,
title = {Conditional collusion 1},
title style = {align=center, font=\small},
xmajorgrids = true,
ymajorgrids = true,
symbolic x coords = {A,B,C,D,E,F},
xtick = data,
xlabel = {Groups},
xlabel style = {align=center, font=\small},
ymin = 0,
scaled y ticks = false,
ylabel = {Probability in \%},
ylabel style = {align=center, font=\small},
yticklabel style = {/pgf/number format/fixed,/pgf/number format/precision=5},
ytick style = {draw = none},
]
\addplot [blue, only marks, mark = square, very thick] coordinates{
(A,0.949)
(B,0.3166)
(C,0.6624)
(D,0.2736)
(E,0.356)
(F,0.6483)
};
\addplot [red, only marks, mark = star, thick] coordinates{
(A,1.0304)
(B,0.337)
(C,0.7281)
(D,0.4717)
(E,0.6288)
(F,0.6866)
};
\addplot [ForestGreen, only marks, mark = triangle, very thick] coordinates{
(A,1.0453)
(B,0.4761)
(C,0.8887)
(D,0.6353)
(E,0.6536)
(F,0.7381)
};
\end{axis}

\begin{axis}[
at = {(axis3.south east)},
xshift = 0.12\textwidth,
width = 0.49\textwidth, 
height = 0.4\textwidth,
title = {Conditional collusion 2},
title style = {align=center, font=\small},
xmajorgrids = true,
ymajorgrids = true,
symbolic x coords = {A,B,C,D,E,F},
xtick = data,
xlabel = {Groups},
xlabel style = {align=center, font=\small},
ymin = 0,
scaled y ticks = false,
ylabel = {Probability in \%},
ylabel style = {align=center, font=\small},
yticklabel style = {/pgf/number format/fixed,/pgf/number format/precision=5},
ytick style = {draw = none},
]
\addplot [blue, only marks, mark = square, very thick] coordinates{
(A,5.3657)
(B,4.1724)
(C,6.2325)
(D,4.0627)
(E,4.2545)
(F,8.1972)
};
\addplot [red, only marks, mark = star, thick] coordinates{
(A,5.8099)
(B,4.544)
(C,6.8094)
(D,9.509)
(E,9.7119)
(F,8.4459)
};
\addplot [ForestGreen, only marks, mark = triangle, very thick] coordinates{
(A,5.8911)
(B,5.1977)
(C,7.7272)
(D,9.3118)
(E,9.7386)
(F,8.8466)
};
\end{axis}
\end{tikzpicture}

\vspace{0.25cm}
\begin{tikzpicture}
\begin{axis}[
name = axis5,
width = 0.49\textwidth, 
height = 0.4\textwidth,
title = {Conditional collusion 3},
title style = {align=center, font=\small},
xmajorgrids = true,
ymajorgrids = true,
symbolic x coords = {A,B,C,D,E,F},
xtick = data,
xlabel = {Groups},
xlabel style = {align=center, font=\small},
ymin = 0,
scaled y ticks = false,
ylabel = {Probability in \%},
ylabel style = {align=center, font=\small},
yticklabel style = {/pgf/number format/fixed,/pgf/number format/precision=5},
ytick style = {draw = none},
legend style = {font=\small,at={(0.35,-0.3)},anchor=north west,legend columns=3},
legend entries = {Schedule S1$\qquad \qquad$, Schedule S2$\qquad \qquad$, Schedule S3}
]
\addplot [blue, only marks, mark = square, very thick] coordinates{
(A,1.8924)
(B,1.0386)
(C,2.1592)
(D,1.5609)
(E,1.7299)
(F,2.8729)
};
\addplot [red, only marks, mark = star, thick] coordinates{
(A,1.9328)
(B,1.0851)
(C,2.217)
(D,2.1959)
(E,2.2547)
(F,2.8891)
};
\addplot [ForestGreen, only marks, mark = triangle, very thick] coordinates{
(A,1.9184)
(B,0.9735)
(C,2.1801)
(D,1.79)
(E,2.2022)
(F,2.9132)
};
\end{axis}

\begin{axis}[
at = {(axis5.south east)},
xshift = 0.12\textwidth,
width = 0.49\textwidth, 
height = 0.4\textwidth,
title = {Dual collusion},
title style = {align=center, font=\small},
xmajorgrids = true,
ymajorgrids = true,
symbolic x coords = {A,B,C,D,E,F},
xtick = data,
xlabel = {Groups},
xlabel style = {align=center, font=\small},
ymin = 0,
scaled y ticks = false,
ylabel = {Probability in \%},
ylabel style = {align=center, font=\small},
yticklabel style = {/pgf/number format/fixed,/pgf/number format/precision=5},
ytick style = {draw = none},
]
\addplot [blue, only marks, mark = square, very thick] coordinates{
(A,3.5948)
(B,2.4046)
(C,4.3766)
(D,3.1588)
(E,3.4004)
(F,5.9763)
};
\addplot [red, only marks, mark = star, thick] coordinates{
(A,3.6913)
(B,2.5298)
(C,4.5943)
(D,5.3445)
(E,5.334)
(F,6.0945)
};
\addplot [ForestGreen, only marks, mark = triangle, very thick] coordinates{
(A,3.7357)
(B,2.4271)
(C,4.7245)
(D,4.5824)
(E,5.266)
(F,6.1465)
};
\end{axis}
\end{tikzpicture}

\caption{The probability of match-fixing opportunities in the last round}
\label{Fig1}

\end{figure}
